\documentclass{article}

\usepackage{arxiv}

\usepackage[utf8]{inputenc} 
\usepackage[T1]{fontenc}    
\usepackage{hyperref}       
\usepackage{url}            
\usepackage{booktabs}       
\usepackage{amsfonts}       
\usepackage{nicefrac}       
\usepackage{microtype}      
\usepackage{lipsum}
\usepackage{graphicx}
\graphicspath{ {./images/} }
\usepackage{amsmath}

\title{Personalized Head-Related Transfer Function Prediction Based on Spatial Grouping}

\author{
Keng-Wei Chang \\
  Department of Electronics and Electrical Engineering \\ National Yang Ming Chiao Tung University \\
  Taiwan\\
  \texttt{alex0976296586@gmail.com} \\
   \And
 Yih-Liang Shen \\
  Department of Electronics and Electrical Engineering \\ National Yang Ming Chiao Tung University \\
  Taiwan\\
  \texttt{yihliang.ee06@nycu.edu.tw} \\
  \And
 Tai-Shi Chi \\
   Department of Electronics and Electrical Engineering \\ National Yang Ming Chiao Tung University \\
  Taiwan\\
  \texttt{tschi@nycu.edu.tw} \\
}

\begin{document}
\maketitle
\begin{abstract}
The head-related transfer function (HRTF) characterizes the frequency response of the sound traveling path between a specific location and the ear. When it comes to estimating HRTFs by neural network models, angle-specific models greatly outperform global models but demand high computational resources. To balance the computational resource and performance, we propose a method by grouping HRTF data spatially to reduce variance within each subspace. HRTF predicting neural network is then trained for each subspace. Simulation results show the proposed method performs better than global models and angle-specific models by using different grouping strategies at the ipsilateral and contralateral sides.
\end{abstract}


\section{Introduction}

In recent years, the rapid development of techniques for virtual reality (VR) and augmented reality (AR) has provided users with exceptional immersive experiences. Presenting highly realistic visuals to users is important, as well as accurately presenting the spatial sound within the virtual environment. The path of the sound from its origin to each of the human ears involves a series of interactions with body parts, including scattering and diffraction with the head, torso, and earlobe. All these interactions contribute to the spatial perception of the sound. Modeling the path as a linear time-invariant system, the impulse response of this system is called the head-related impulse response (HRIR). The frequency domain representation of the HRIR is the head-related transfer function (HRTF). The HRTF integrates binaural cues (ITD, ILD) and spatial filters formed by interactions with the body parts. Knowing individual’s HRTFs allows for effectively synthesizing spatial sound from mono audio. Although the HRTF contains magnitude and phase responses, we only consider the magnitude response of the HRTF in this paper as in most HRTF prediction literature. Therefore, the HRTF in this paper mostly represents the magnitude response of the HRTF.

Various methods have been proposed to measure individual’s HRTFs. Acoustic measurement methods \cite{10.1121/1.412407, majdak2007multiple, app10145014} offer precise results through direct measurements which entail costly and time-intensive processes with specialized equipment, and controlled environments. Alternatively, database matching methods \cite{7471692, 1285855, 1048207, TORRESGALLEGOS201584} and mathematical model simulation methods \cite{10.1121/1.3257598, 625596} were proposed to simulate the HRTFs. Currently, the regression method is widely explored for estimating individual’s HRTFs based on anthropometric parameters. Some studies have introduced innovative approaches, such as principal component analysis (PCA) and deep neural networks (DNNs), to predict personalized HRTFs effectively \cite{7602913, FINK2015162, 10.1121/1.402444}. These advancements hold promise for improving the accessibility and efficiency of personalized spatial audio rendering techniques.

Chun et al. \cite{chun2017deep} introduced a DNN model, which, upon inputting data of the head, torso, and pinnae, autonomously determines feature weights and directly obtains the HRIRs. To overcome the challenges of pinnae measurements in real life, Lee et al. \cite{app8112180} proposed employing convolutional neural networks (CNNs) to extract features from pinna images instead of directly measuring individual pinnae. However, the methods used distinct NN models for different azimuth and elevation angles, resulting in a total of 1250 models (25 azimuth angles and 50 elevation angles), which limits their practical usability. In contrast, Wang et al. \cite{Wang2021GlobalHP} incorporated spherical harmonic transform and proposed a global HRTF personalization method based on CNNs.


While using angle-specific models to predict HRTFs produces good results, this approach requires a large number of models and lacks the ability to predict unseen angles. Another drawback of this approach is that less data can be used in training each of the angle-specific models, potentially producing an overfitting model. On the other hand, the approach of training a global model for all angles often produces unsatisfactory results due to the large variations between HRTFs at different angles \cite{10.1121/10.0011575,10413449}. To balance these two approaches, we propose a personalized HRTF prediction method based on spatial grouping. The idea of our method is to group HRTFs with similar characteristics for reducing HRTF variations within each group. In each of the four groups, a neural network model, containing a variational autoencoder (VAE) and a DNN, is trained to predict intra-group HRTFs. The proposed method significantly reduces the number of models required to estimate all angles compared with the angle-specific model approach, and achieves large improvement in prediction accuracy compared with the global model approach in terms of the log spectral distance (LSD) measure. 
Experiment results demonstrate the proposed method can produce comparable LSD scores with the angle-specific model. 
The main contributions of this paper are summarized below:
\begin{enumerate}
  \item We propose the idea of grouping HRTFs spatially and build different NN models to predict intra-group HRTFs.
  \item We investigate spatial grouping strategies and show that ipsilateral and contralateral ears need different grouping rules to produce high LSD scores.
  \item We propose a hybrid method, which well balances the number of NN models and prediction accuracy, for practical HRTF personalization.
\end{enumerate}

The rest of the paper is organized as follows. In Section 2, we introduce the dataset and related research. In Section 3, the proposed method and experimental settings are presented. Experiment results and discussion are given in Section 4. Finally, we conclude this paper in Section 5.

\section{Background}
\subsection{Database}

In our experiments, we used the CIPIC database \cite{969552}, which offers high spatial resolution measurement data of HRTFs and is used in many HRTF-related studies. Furthermore, this database is often used to explore the relationship between anthropometric features and HRTF \cite{10.1121/10.0004128, ghorbal2017pinna}. The database comprises HRTF data from 45 subjects, including measurements from two different-sized KEMAR heads \cite{10.1121/1.412407}. For each subject, it consists of 20 measurements of the pinnae (10 measurements for each ear) and 17 measurements of the main torso, totaling 37 anthropometric parameters. Additionally, it contains HRIRs for each ear in 1250 directions. The length of HRIR data for each direction is 200 sample points, sampled at a frequency of 44.1 kHz, resulting in a length of approximately 4.5 ms. Moreover, the data have been compensated for factors such as reflections in the measurement space and recording equipment gains.

\begin{figure}[th]
  \centering
  \includegraphics[width=0.65\linewidth]{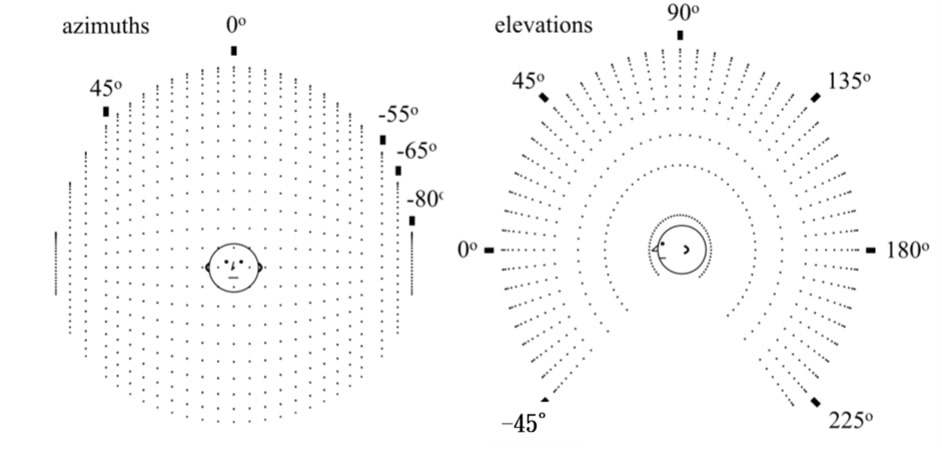}
  \caption{Coordinate point distribution in CIPIC: (a) frontal view (b) lateral view \cite{sridhar2015capturing}.}
  \label{fig:Coordinate point distribution}
\end{figure}

The interaural-polar coordinate system is used in CIPIC. The measured azimuth angles are at -80°, -65°, -55°, -45°, then every 5° from -45° to +45°, and at +55°, +65°, +80°, for a total of 25 azimuth angles. The elevation angles are from -45° to 230.625° in a step of (360°/64), resulting in 50 elevation angles. Consequently, there are 1250 directions for each ear of each subject in the database. Fig. \ref{fig:Coordinate point distribution} shows the spatial distribution of these 1250 directions. However, some subjects have incomplete measurements, the total number of subjects with complete 1250 HRTF data is 35.

\subsection{Prediction methods using autoencoder (AE) and VAE}

In \cite{8683814}, authors proposed a HRTF prediction method by firstly compressing the HRTFs into latent codes using an AE, aiming to alleviate overfitting arised from limited data. Then, a separate DNN leverages the anthropometric parameters of the user as input features to predict the latent codes. Finally, the decoder of the AE reconstructs the corresponding HRTFs. In summary, the method works by inputting the anthropometric parameters of the user into a DNN to predict latent codes, which are then decoded to construct the HRTFs.

In \cite{10.1121/10.0011575}, the proposed method differs by employing a VAE instead of a standard AE. Unlike the AE, which maps data points to specific latent codes, the VAE models the distribution of latent codes of the data, allowing for a more nuanced capture of characteristics of continuous data. Consequently, the HRTFs are first encoded to their corresponding latent codes using VAE, while the anthropometric parameters are encoded to their respective latent codes using an AE. Subsequently, a DNN predicts the latent codes of HRTFs corresponding to angles by inputting the latent codes of anthropometric parameters and then decodes the predicted latent codes back to the HRTFs via the VAE decoder. This method enhances the model's ability to represent the underlying distribution of the data and can predict HRTFs at unseen angles.

\section{Proposed Method}
\subsection{Data preprocessing}
\subsubsection{Anthropometric parameters}

In the CIPIC database, each subject has 37 anthropometric parameters. Although each ear is associated with 1250 HRTFs, we focus on HRTFs of the left ear in this paper without loss of generality. Therefore, we discard 10 parameters of the right ear and use only 27 relevant parameters for our study. These parameters are normalized using the following equation as in \cite{chun2017deep}:
\begin{align}
  f'_i &= (1+\exp(-\frac{f_i-\mu_i}{\sigma_i}))^{-1}
  \label{equation:eq1}
\end{align}
where $f_i$ indicates the $i$-th anthropometric parameter of the 27-dimensional data; $\mu$ and $\sigma$ are the mean and standard deviation across subjects.

After obtaining the normalized parameters, we combine the 3-dimensional spatial information with this 27-dimensional data to have the complete 30-dimensional input data. The 3-dimensional spatial information comprises three Cartesian coordinates of the direction. Since the CIPIC database use the interaural-polar coordinate system, we convert the interaural-polar coordinates into Cartesian coordinates using the following formula:
\begin{align}
  (x,y,z) = (R\cos\theta\cos\phi, R\sin\theta, R\cos\theta\sin\phi)
  \label{equation:eq2}
\end{align}
where $R$ is the radius; $\theta$ and $\phi$ are the azimuth and elevation angles.

\subsubsection{HRTFs}
The CIPIC database contains the original HRIRs of each subject. We transform HRIRs into HRTFs by firstly performing a 512-point discrete Fourier transform (DFT), then using a constant Q filter bank (Q=8) for smoothing, and finally computing the logarithm of the magnitude responses to represent the HRTFs in decibels. Subsequently, we retain only the frequency range from 200 Hz to 15000 Hz as in the study \cite{7378943}.
Upon extracting this frequency spectrum, we convert the original linear frequency bands into logarithmic frequency bands. Therefore, each of the HRTFs utilized in our study is represented as a 173-dimensional vector. Additionally, we apply min-max normalization to the processed logarithmic magnitude responses to confine the values within the range of 0 and 1.

\begin{figure}[th]
  \centering
  \includegraphics[width=0.6\linewidth]{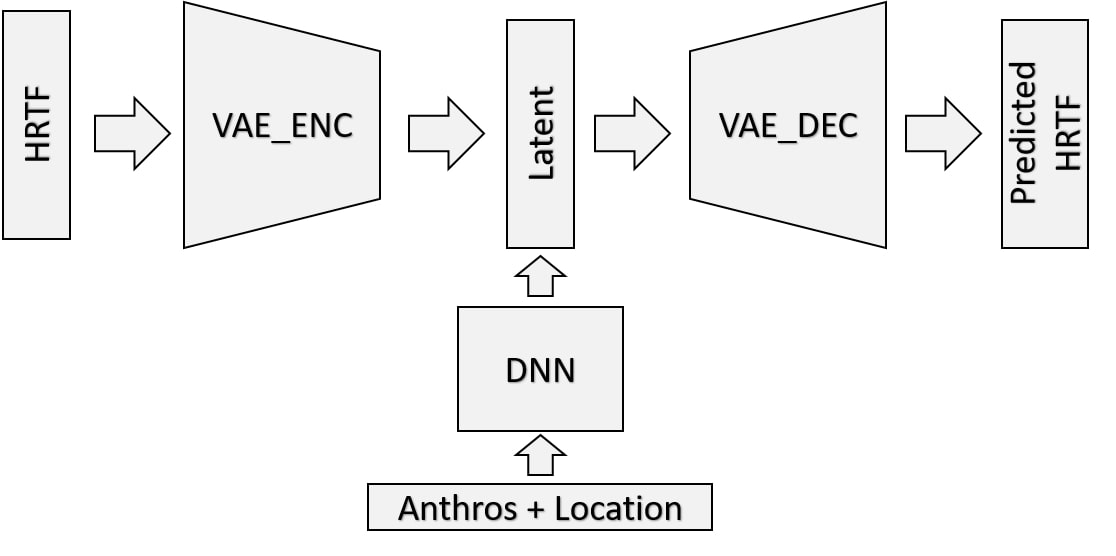}
  \caption{Basic structure of the proposed model.}
  \label{fig:basic model}
\end{figure}

\subsection{Model structure}
The model's basic architecture, depicted in Fig. \ref{fig:basic model}, comprises two major components, i.e., a VAE and a DNN. We adopt the concatenation architecture of an AE and a DNN from \cite{8683814}, and replace the AE with a VAE \cite{10.1121/10.0011575} to predict HRTFs for unseen angles. As described in \cite{10.1121/10.0011575}, the VAE introduces a probability distribution for data such that it can effectively capture continuous variations within data. Moreover, owing to the continuous nature of the VAE, its generative capacity is also continuous, rendering the latent variables more interpretable and manipulatable in practical applications than AE. The VAE is divided into an encoder (VAE\_ENC) and a decoder (VAE\_DEC). Both modules have three layers: input, hidden, and output layers. The encoder's layers are with 128, 64, and 32 neurons, while the decoder possesses a reverse configuration. All layers except the final output layer employ the ReLU activation function and batch normalization. The final output layer only uses a Sigmoid activation function to deliver the predicted HRTFs. In the VAE\_ENC, two additional layers are used to generate 32-dimensional latent codes for the mean and variance of the Gaussian distribution. One layer outputs the mean, while the other layer outputs the variance.

The DNN takes in 30-dimensional 'Anthros+Location' data, which is a combination of 27-dimensional anthropometric parameters and 3-dimensional Cartesian coordinates of the spatial location, to predict the 32-dimensional latent codes of the Gaussian distribution. It consists of a shared input layer, a shared hidden layer, and two separate modules for mean and variance, each of which has two hidden layers and an output layer. This setup creates a split structure after the shared 64-neuron hidden layer. Each of the succeeding two hidden layers has 128 neurons, and the final output layer has 32 neurons. Similar to the VAE, the DNN uses the ReLU activation function and batch normalization across all layers except the two final output layers.

\subsection{Training strategy}
When training the VAE, we used the mean square error (MSE) between the logarithmic magnitude responses of the reconstructed HRTF and the original HRTF as the loss function. We utilized the Adam optimizer with a learning rate of 0.00001 to update the parameters of the VAE. When training the DNN, the loss function comprises the MSE of the predicted latent code, and the LSD between the logarithmic magnitude responses of the reconstructed HRTF and the original HRTF. We utilized the Adam optimizer with a learning rate of 0.0001 to optimize the parameters of the DNN. Note that the parameters of VAE were fixed for training the DNN.

Because only limited data was available, we performed leave-one-out cross-validation in our experiments to test the robustness and reliability of our model. In each trial, HRTFs of 34 subjects, out of 35 subjects, were used for training, and HRTFs of the remaining subject were reserved for evaluation. Since we only considered HRTFs of the left ear, the dataset contained 43,750 $(35 \times 1250)$ left-ear HRTFs in total. Additionally, to assess the model's prediction performance on unseen angles, we excluded 20\% of the locations from training and used them as unseen data for evaluation. In summary, in each trial, there were 34,000 HRTFs for training $(34 \times 1250 \times 0.8)$, 1000 HRTFs for seen angle evaluation $(1 \times 1250 \times 0.8)$, and the remaining 250 HRTFs for unseen angle evaluation $(1 \times 1250 \times 0.2)$.

\section{Experiments and Results}

\subsection{Evaluation metric: LSD}
In the task of predicting personalized HRTFs, LSD is mostly used as the objective evaluation metric. Based on the score, we could know whether the predicted HRTF is similar to the ground truth or not. LSD is calculated as follows:
\begin{align}
  LSD(H,\hat{H}) = \sqrt{\frac{1}{k_2-k_1+1}\sum_{k=k_1}^{k_2}(20\log_{10}(\left\lvert\frac{H(k)}{\hat{H}(k)}\right\rvert))^2}
  \label{equation:eq1}
\end{align}
where $H(k)$ and $\hat{H}(k)$ are the magnitude responses of the original HRTF and the predicted HRTF, and $k$ is the frequency bin index.

\subsection{Experiment 1: grouping by spatial location (SL)}
Our purpose of reducing the variance between the model's input data can be fulfilled by finding a reasonable boundary to divide the global space into subspaces. The goal was to reduce the intra-subspace variations such that training of the model becomes easier.

In the first set of experiments, we divided the original global space into four subspaces, left-front, left-back, right-front, and right-back subspaces, based on the spatial location. The left-front subspace encompassed azimuth angles from -80° to 0° and elevation angles from -45° to 90°, while the left-back subspace covered azimuth angles from -80° to 0° and elevation angles from 90° to 230.625°. Conversely, the right-front space spaned azimuth angles from 0° to 80° and elevation angles from -45° to 90°, and the right-back subspace included azimuth angles from 0° to 80° and elevation angles from 90° to 230.625°. This grouping strategy was referred to as Spatial Location (SL). Consequently, four basic models were trained to predict HRTFs within the four subspaces.

Table \ref{tab:result1} presents the results using the SL grouping strategy in predicting both seen and unseen angles, alongside comparisons with results shown in prior studies \cite{8683814, 10.1121/10.0011575, 10413449}. Not surprisingly, the results demonstrate that our four-subspace method using SL grouping strategy outperforms the global models proposed in \cite{10.1121/10.0011575,10413449} in predicting HRTFs at both seen and unseen angles. While the model proposed in \cite{8683814} achieves higher scores on seen angles, it's crucial to note that it's an angle-specific model, requiring a large number of models to predict HRTFs at all angles. Our four-subspace method offers a good balance between performance and system complexity.

\begin{table}[ht]
\caption{Mean LSD (in dB) across all locations of the proposed method using the SL grouping strategy and other compared methods for seen and unseen angles}
  \label{tab:result1}

\vskip3pt
\centering
\begin{tabular}{ccc}
\hline\hline
    &  \textbf{seen angles} & \textbf{unseen angles}\\

\hline
\hline
        SL &  3.59 & 3.56\\
         \hline
        \textbf{\cite{8683814}}& 3.43 & $-$\\
         \hline
        \textbf{\cite{10.1121/10.0011575}} & 4.15 & 4.29\\
         \hline
        \textbf{\cite{10413449}} & 4.17 & $-$\\
\hline\hline
\end{tabular}
\end{table}

\begin{figure}[th]
  \centering
  \includegraphics[width=0.6\linewidth]{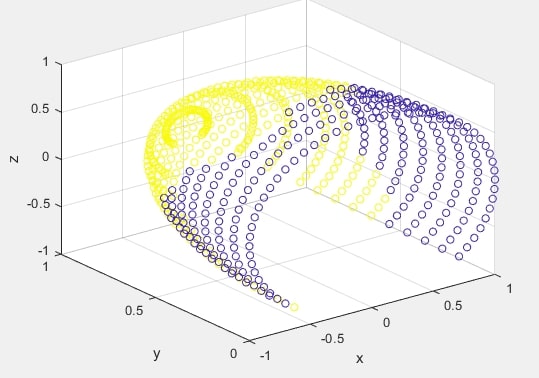}
  \caption{Grouping results on contralateral angles to the left ear based on the diffraction effect. The head is at the location of $(x, y, z)$ = (0, 0, 0) and facing the y-axis. Yellow and blue locations divided by using the DE strategy clearly form two groups, referred to as the inner and outer groups.}
  \label{fig:diff_group}
\end{figure}

\subsection{Experiment 2: grouping by diffraction effect (DE)}
Further analysis showed that the proposed method with the SL strategy performed much better on the ipsilateral angles than on the contralateral angles as demonstrated in Table \ref{tab:result2}. HRTFs of the contralateral side are heavily influenced by the diffraction effect due to head shadowing \cite{cheng1999introduction}. In other words, the low-frequency parts of HRTFs at specific angles on the contralateral side exhibit significant changes due to diffraction effects. Leveraging this fact, we proposed another grouping strategy for the HRTFs on the contralateral side. 
By calculating the average energy of the HRTFs in the low-frequency band (0.2 kHz-0.5 kHz) of 35 subjects, we adopted a threshold of 0.5 to divide the HRTFs into two groups. This grouping strategy was referred to as Diffraction Effect (DE).

Fig. \ref{fig:diff_group} shows the grouping result (yellow: inner group; blue: outer group). We can observe that the yellow locations, whose HRTFs have higher energy at low frequencies, are mainly the locations blocked by the head on the contralateral side. It is also consistent with findings in \cite{shaw1974external} that the diffraction effect mainly occurs near the locations blocked by the head. The circular diffraction zone near the contralateral side (+90° azimuth angle) is blocked by the head and is called the Bright Spot \cite{shaw1974external}. Table \ref{tab:result2} showcases that the performance on the contralateral side by using the DE strategy (seen: 3.29, unseen: 3.32) is much better than by using the SL strategy (seen: 3.94, unseen: 3.90). One-way ANOVA tests under both seen ($F(1,36398)=2597.55,\ p<0.05$) and unseen ($F(1,9098)=527.18,\ p<0.05$) conditions show the differences are statistically significant.
Furthermore, using the DE strategy on the contralateral side for seen angles (seen: 3.29) approaches the performance on the ipsilateral side using the SL strategy (seen: 3.24), and the one-way ANOVA test ($F(1,36398)=1.1,\ p=0.2932$) indicates no significant difference. Note that, on the ipsilateral side, the performance by using the DE strategy (seen: 3.42, unseen 3.41), not shown in the table, is worse than by using the SL strategy (seen: 3.24, unseen: 3.22).


\begin{table}[th]
  \caption{Mean LSD on Ipsilateral and Contralateral sides using different grouping strategies}
  \label{tab:result2}
  \centering
  \begin{tabular}{ |c|c|c|c|c|c| }
    \hline
    \hline
    \multicolumn{2}{|c|}{\textbf{SL Ipsilateral}} 
                                    &\multicolumn{2}{c|}{\textbf{SL Contralateral}}
                                    &\multicolumn{2}{c|}{\textbf{DE Contralateral}}\\
    \hline                                    
    \textbf{\ seen\ } & \textbf{\ unseen\ } &\textbf{\ seen\ } & \textbf{\ unseen\ } &\textbf{\ seen\ } & \textbf{\ unseen\ }\\  
    \hline
    $3.24$ & $3.22$ & $3.94$ & $3.90$ & $3.29$ & $3.32$ \\
    \hline
    \hline
  \end{tabular}
\end{table}

\subsection{Experiment 3: hybrid grouping}
\begin{figure}[th]
  \centering
  \includegraphics[width=0.65\linewidth]{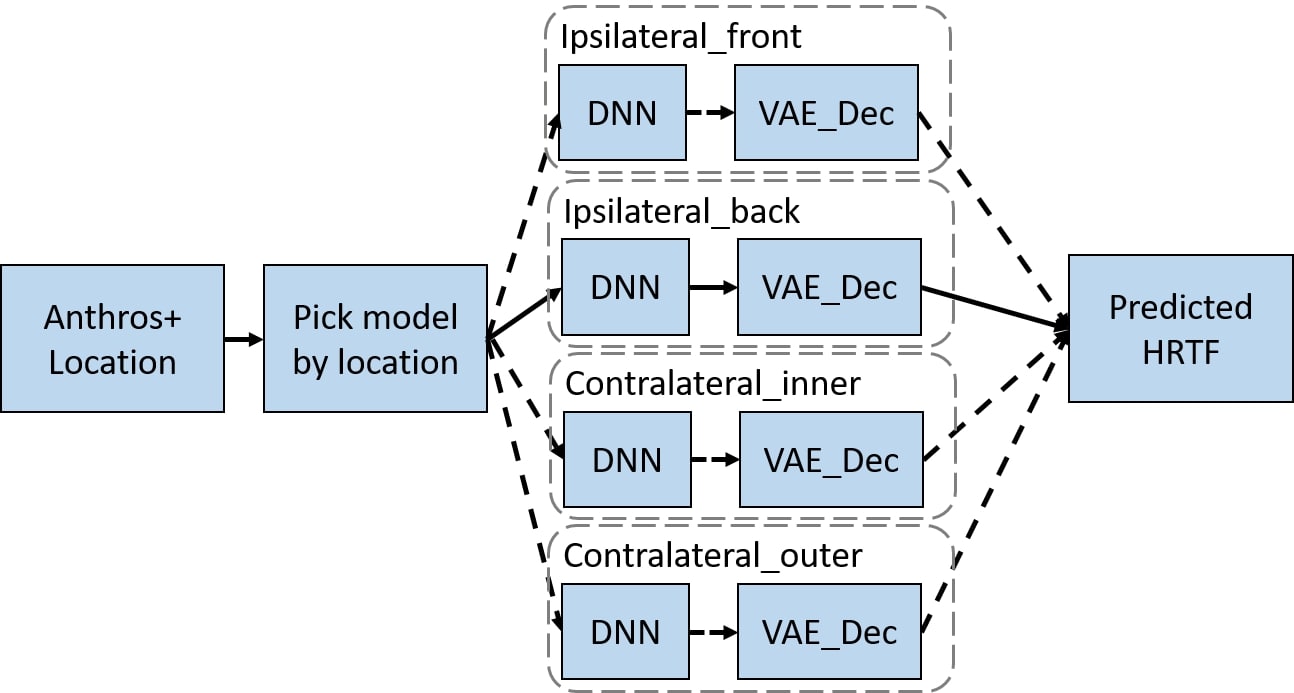}
  \caption{Architecture of the hybrid-grouping method. It picks up the specific model based on the location to predict the HRTF.}
  \label{fig:hybrid}
\end{figure}

Table \ref{tab:result2} shows that the SL and the DE grouping strategies are respectively suitable for the ipsilateral and the contralateral sides. Therefore, to have the optimal performance, we proposed a hybrid method for predicting HRTFs by using the SL strategy for ipsilateral angles and the DE strategy for contralateral angles. The architecture of the hybrid method is shown in Fig. \ref{fig:hybrid}, and
the corresponding results are shown in Table \ref{tab:result3}.

\begin{table}[th]
  \caption{Mean LSD across all locations using the SL and the hybrid grouping strategies}
  \label{tab:result3}
  \centering
  \begin{tabular}{ |c|c|c|c|c|c| }
    \hline
    \hline
    \multicolumn{2}{|c|}{\textbf{SL Grouping}} 
                                    &\multicolumn{2}{|c|}{\textbf{Hybrid Grouping}}
                                    &\multicolumn{2}{|c|}{\textbf{\cite{8683814}}}\\
    \hline
    \textbf{ seen } & \textbf{ unseen } & \textbf{ seen } & \textbf{ unseen } & \textbf{ seen } & \textbf{ unseen } \\  
    \hline
    $3.59$ & $3.56$ & $3.26$ & $3.27$ & $3.43$ & $-$ \\
    \hline
    \hline
  \end{tabular}
\end{table}

The results in Table \ref{tab:result3} demonstrate that using the hybrid strategy achieves the highest performance for both seen and unseen angles. Surprisingly, it even outperforms the angle-specific model in \cite{8683814}. We attribute this success to the grouping of HRTFs with similar characteristics, which effectively increases the data volume (compared with angle-specific models) and reduces data variance (compared with global models) for model training. It allows the model to learn more robust representations, leading to improved performance in predicting personalized HRTFs.

\section{Conclusions and Future work}

In this study, we propose a VAE based method for predicting personalized HRTFs to enhance spatial audio experience. To reduce data variance for model training, we propose different grouping strategies, namely the SL and the DE strategies, to effectively partition the HRTF data into subspaces based on spatial and diffraction characteristics. By leveraging the grouping strategies, we train separate models for 4 subspaces using VAEs and DNNs, resulting in improved accuracy in predicting personalized HRTFs.

Furthermore, we implement a hybrid-grouping method that combines both SL and DE strategies, and demonstrate its superior performance compared with global and angle-specific models. The hybrid-grouping method uses the SL strategy for ipsilateral angles and the DE strategy for contralateral angles. It effectively leverages similarities in HRTF characteristics to increase data volume and reduce variance for model training, resulting in enhanced prediction accuracy for both seen and unseen angles. In future, we will verify its superior performance using subjective listening tests. Moreover, we divide the global space into 4 subspaces in this paper. The optimal number of subspaces to balance the performance and system complexity, and the optimal criterion instead of the average energy in the low frequency band for DE strategy are questions worth exploring in future. Meanwhile, we will evaluate the proposed grouping strategy on other personalization frameworks such as \cite{Wang2021GlobalHP,
bhattacharya2020optimization}.

\bibliographystyle{IEEEtran}
\bibliography{refs}
\end{document}